# The magneto-Leidenfrost effect in ferrofluid droplets

**Abhishek Kumar Jaiswal**[#], **Neeladri Sekhar Bera,** and **Purbarun Dhar**[*]

Hydrodynamics and Thermal Multiphysics Lab (HTML), Department of Mechanical Engineering, Indian Institute of Technology Kharagpur, West Bengal–721302, India

[#]E–mail: akjaiswal.stem@gmail.com

*Corresponding author: E–mail: purbarun@mech.iitkgp.ac.in

## ABSTRACT

***Hypothesis***: The dynamic Leidenfrost effect (LFE) and behaviour of impinging colloidal droplets is strongly influenced by the impact and spreading paradigms. LFE-actuated rebound and levitation occurs due to enhanced spreading and near-frictionless recoil over the intervening vapour layer, providing opportunities for external-field-stimulus-aided modulation and control of impact outcomes, and the resulting boiling-LFE behaviour.

***Methods***: Magnetic field modulated LFE onset, dynamics and boiling transport of stable aqueous nano-$Fe_2O_3$ based ferrofluid droplets was studied using high-speed-imaging. The interplay between magnetic, inertia, and visco-capillary forces on droplet spreading, magneto-LFE-driven rebound conditions, residence time, and post-impact regimes was analysed using dimensionless parameters: maximum spread factor, $\beta_{\max}$, Weber number, $We_o$, and magnetic Bond number, $Bo_m$.

***Findings***: We report a purely new phenomenon, namely magneto-Leidenfrost effect (MLFE), wherein magnetic field induces LFE-aided onset of droplet rebound at substrate temperatures ($T_s$) below the zero-field dynamic Leidenfrost temperature (LFT). The critical $Bo_m$ for the onset of MLFE decreases with increasing $We_o$. Increasing the nanoparticle concentration permits the onset even at considerably lower $We_o$. At elevated $T_s$, the residence time is noted as $Bo_m$ dependent. At much higher $T_s$, increasing $Bo_m$ promotes formation of radial filamentous structures, leading to complete droplet fragmentation. We also propose a theoretical framework that explains magnetic field-driven spreading enhancement and rebound, and predicts $\beta_{\max}$ of MLFE droplets in agreement with experiments. Our findings provide valuable insights into the novel realm of field dictated LFE, and hold significant implications towards the design of frictionless, rapid colloid droplet transport systems, and targeted droplet manipulation/activation for advanced thermal management.

## 1. Introduction

Impact or interactions of a droplet on a heated surface, at temperatures significantly above the boiling point of the liquid, culminates in rapid evaporation generated vapour layer between the droplet and the substrate. This intervening vapour layer, often micron-level thick, prevents direct solid-liquid contact, and acts as a near-frictionless cushion that enables the droplet to levitate, hover, or propel above the surface. The vapour layer behaves as an insulating film, diminishing the heat transfer, and thereby resulting in a significant increase in phase transition time. First observed by a physician, J. G. Leidenfrost [1] in 1756, it is known as the Leidenfrost effect (LFE). The LFE exhibits a dualistic paradigm. On one hand, it is highly undesired regime to heat transfer-based design considerations in a myriad of applications, such as spray cooling in fire extinguishment, and electronics cooling systems [2,3], material quenching [4], cooling of nuclear fuel rods [5], and fuel drop impingement in internal-combustion engines [6]. On the contrary, it provides opportunities for controlling preferential motion and self-propulsion of droplets [7-9] in advanced thermofluidic devices. For gently deposited droplets, the Leidenfrost temperature (LFT, $T_L$) corresponds to the maximum droplet evaporation time. However, for impinging droplets, the corresponding dynamic Leidenfrost temperature ($T_{DL}$) is identified as the lowest surface temperature at which onset of droplet rebound occurs with minimal spraying or droplet disintegration [10-13].

Existing theoretical works include investigations on shape [14] and geometry [15] of Leidenfrost droplets, and the underlying vapour layer, and models predicting Leidenfrost temperature [14,16-19]. The Leidenfrost temperature is determined by balancing dynamic pressure due to droplet impact with the vapour pressure supporting the droplet levitation.

Using lubrication approximation, the scale of vapour pressure is obtained by modeling the vapour flow through the thin space between droplet and substrate. This can be related to vapour generation rate at the liquid-vapour interface by incorporating relevant length and time scales of spreading Leidenfrost droplets [10,20-22]. By scaling time for vapour layer growth during the early inertia-dominated spreading stage, Park and Kim [23,24] established a relation between vapour layer thickness and substrate superheat. Combining this with an energy-balance-based estimate of the maximum spread diameter, they derived a scale of Leidenfrost temperature considering instantaneous vapour layer formation after impact. They showed that the Leidenfrost temperature increases with the impact velocity.

Researchers have attempted to modulate the $T_{DL}$ by changing the droplet composition via addition of polymers [25,26], surfactants [27,28], and nanoparticles [29,30]. Polymer additives were found to supress droplet rebound through the formation of elastic polymer chains, while surfactants increased $T_{DL}$ by reducing surface tension, and suppressing droplet coalescence. Prasad et al. [30] reported an increase in $T_{DL}$ due to dispersion of a small amount of nanoparticles in the liquid; attributing it to the formation of residual deposits that alter bubble coalescence and vapour-film stability. They also provided a boiling regime map for nanocolloids, and noted that unlike water droplets, nanocolloids exhibit spraying behaviour. Researchers have extensively studied the effect of impact velocity on the $T_{DL}$ [23,31-33] and found that at higher Weber numbers, $We_o$ (ratio of inertia to surface tension forces), a thicker vapour layer is essential to sustain the larger impact momentum, requiring higher surface temperatures for rebound. In contrast, many recent studies [28,30,34-37] report a decrease in $T_{DL}$ with $We_o$. They attribute this behaviour to the enhanced heat transfer through the larger contact spread-area at higher impact velocities [28,30,37].

In the recent past, surface modification-based approaches have been explored towards control over LFT. Micro-textured surfaces, which behave as micropillars and penetrate through the vapour layer, thereby suppressing droplet levitation by rewetting the surface [17,38,39] have been studied. For higher pillar density, however, the increased capillary wetting ceases to dominate and de-wetting vapour pressure increases due to enlarged heat transfer area and enhanced flow resistance to escaping vapour, resulting in decrease in the Leidenfrost temperature [40,41]. The highly mobile nature of Leidenfrost droplets has motivated extensive efforts towards their controlled and directional transport [7-9,42,43]. Linke et al. [7] and Lagubeau et al. [8] demonstrated that Leidenfrost droplets and levitating dry-ice blocks, respectively, can self-propel on asymmetric ratchet surfaces due to vapour-driven forces arising from the directional rectification of anisotropic vapour flow. Li et al. [9] used collective interactions between planar ratchet arrays to achieve enhanced manoeuvrability and control over Leidenfrost droplet transport.

Further, non-tactile stimulus-based methods [44-48] have proven effective for droplet actuation and control. Few works [49-51] have employed electric fields between Leidenfrost droplets and the substrate to suppress the droplet levitation and enhance heat transfer by attracting droplets towards substrate using electrostatic forces. A handful studies [52,53] have explored the use of magnetic fields to trap Leidenfrost droplets and modulate their evaporation rates. While these studies partially address the practical limitations of micro/nano-textured and ratcheted surfaces, such as complex and costly fabrication, susceptibility to fouling, and limited scalability, they focus primarily on Levitation suppression and are largely restricted to near-static droplets ($We_o \approx 0$).

In our novel approach, we employ magnetic field to alter the very behaviour of impinging ferrofluid Leidenfrost droplets, a study that remains completely unexplored in the literature. Unlike previous studies that manipulate Leidenfrost droplets through surface micro-texturing [7,40], the present method is based on a contact-free external stimulus and does not require permanent modification of the substrate surface. Our research reveals that a carefully tailored combination of vertical and horizontal magnetic forces, applied via a non-uniform magnetic field environment, can trigger Leidenfrost rebound under conditions where droplets would otherwise remain grounded. We investigate thoroughly the associated ferro-hydrodynamics, and their influence on magnetic-Leidenfrost effect (MLFE)-driven rebound conditions. Supported by a theoretical model, we systematically examine the role of magnetic forces during different stages of droplet impact and spreading, and show how magnetic fields can be used to control impact outcomes for MLFE, over a wide range of impact conditions, nanoparticle concentrations, and substrate superheats indicating the broad applicability of the method. We further demonstrate control over MLFE droplet residence time and the size of daughter droplets produced during fragmentation by appropriately tuning the magnetic field strength and substrate temperature. The findings reveal a new phenomenon of the MLFE in ferrofluids, offer a potential route for frictionless and rapid droplet transport, targeted heat-transfer modulation in thermal management applications, controlled reagent transport and deposition, microdroplets sorting, and inkjet printing technologies.

## 2. Materials and methodologies

The schematic of our experimental setup is illustrated in Fig. 1. We used a smooth Aluminium plate (25 mm×25 mm) as the test substrate. $T_s$ in the range of 150 °C to 225 °C were maintained using a precision, digitized plate-heater (Holmarc Opto-Mechatronics Ltd., India), with an accuracy of ± 2 °C and integrated T-type thermocouple based control. The magnetic field was generated between the horizontal poles of an electromagnet (Holmarc Opto-Mechatronics Ltd., India) by supplying current from a DC power source. A GaAs sensor-based Hall effect Gauss meter (Holmarc Opto-Mechatronics Ltd., India) was used to measure the magnetic field strength within the whole study space. A rail-mechanism was designed between the heater top-surface and the space between the magnetic poles, such that the superheated substrate was effortlessly slid from the heater within the magnetic field region rapidly during the experiment. A micro-syringe with a volumetric accuracy of ± 0.1 μl, fit with a flat-tipped 22-gauge needle, was attached to a height-adjustable droplet dispensing mechanism (DDM) to dispense droplets from different heights. A high-speed camera (Photron, UK) with 105 mm macro lens (Sigma) was used to capture the droplet interaction events on the hot substrate. The images were recorded at 4000 frames per second, at 1280 × 1024 pixels.

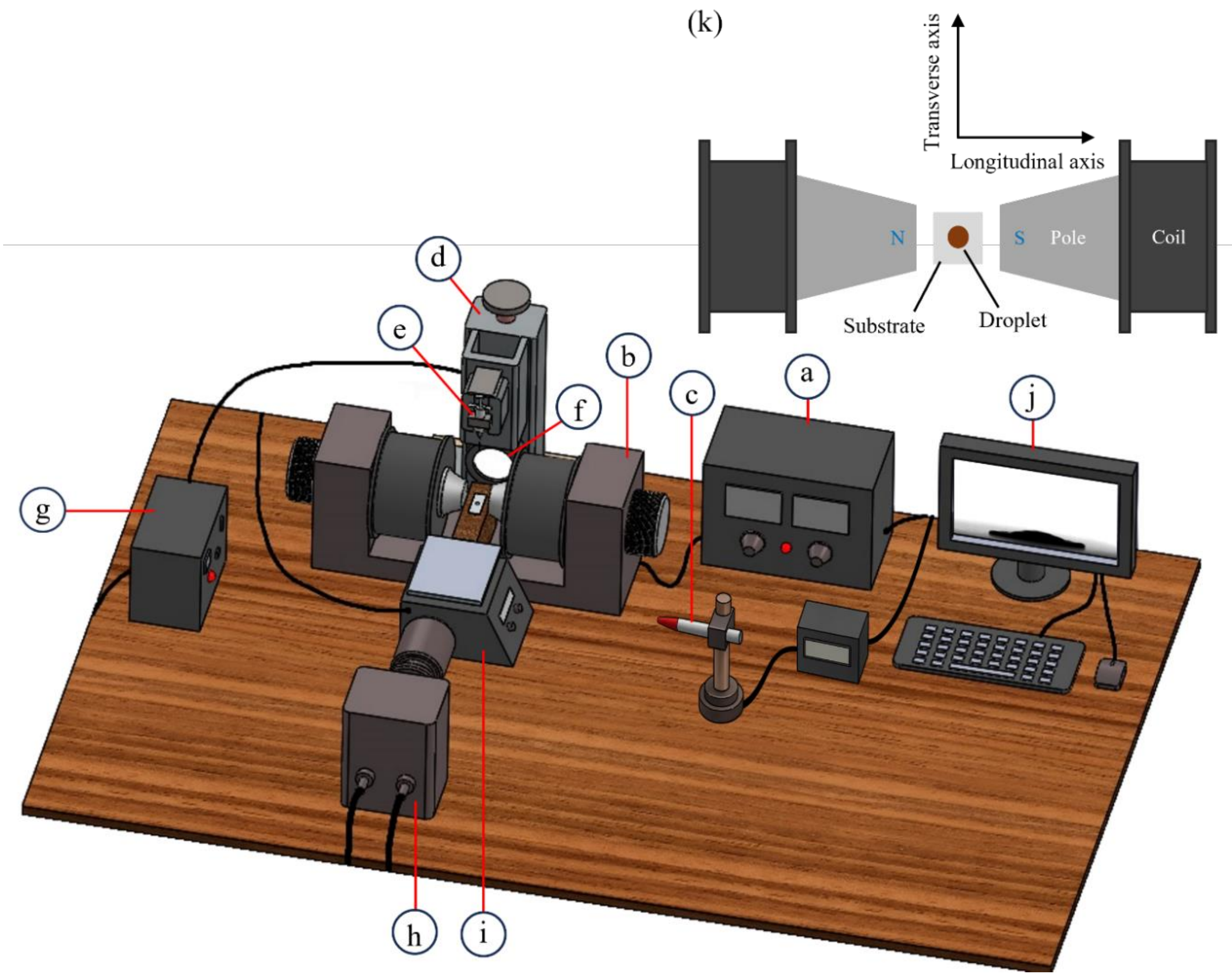

**Fig. 1.** Schematic of the experimental setup: (a) Electromagnet power supply, (b) electromagnet unit, (c) Gauss meter, (d) droplet dispensing mechanism (DDM), (e) micro-syringe, (f) backlight with diffuser screen, (g) DDM and backlight illumination controller, (h) high-speed camera, (i) heater unit, and (j) computer for data acquisition and camera control. Inset: (k) Illustrative top view of the magnetic setup with substrate, post-impact droplet, and reference axes.

Before each experiment, the test surface was cleaned with de-ionized (DI) water, followed by isopropanol to remove contamination. A T-type thermocouple embedded within the substrate was used to monitor the substrate temperature with an accuracy of ± 2 °C. In-house synthesized water-based stable ferrofluids were the test fluids. A colloidal dispersion of $Fe_2O_3$ nanoparticles (30-40 nm, Alfa Aeser, India) were dispersed in DI water to obtain 5% w/w and 7.5% w/w ferrofluids. The magnetic nanoparticles have a coating of citric acid (supplied by the manufacturer) to improve their stability in water. First, the nanoparticles were thoroughly mixed in the DI water by mechanical stirrer for 20-30 minutes. Thereafter, it was ultra-sonicated for 2 hours using probe type sonicator (Oscar ultrasonics, India). The colloidal stability were checked through zeta potential measurements, with the obtained values lying outside the range of -30 mV to +30 mV, indicating adequate stability. Moreover, the synthesized ferrofluids were found to remain stable for durations significantly longer than the experimental timescales (of a few seconds).

The viscosity of the ferrofluids was measured using a rotational rheometer (Anton Paar) with a parallel plate geometry, attached to a magneto-rheology module. Surface tension of the test fluids was measured by the pendant drop method. The static contact angles ($\theta$)on the substrate surface in the presence and absence of magnetic fields were determined through image analysis (ImageJ software) and were found to be ~(70 ± 8)°. The physical properties of the ferrofluids are listed in Table 1. The variation in the properties was within ± 5%, and each set of experiments was repeated three times in the present study.

**Table 1**: Properties of the ferrofluids used in the present work

| **Property** | **5% w/w** | **7.5% w/w** |
|---|---|---|
| Density, $\rho$ (kg/m$^3$) | 1039.56 | 1061.50 |
| Viscosity, $\eta$ (Pa-s) | 0.0363 | 0.0540 |
| Surface tension, $\sigma$ (N/m) | 0.0700 | 0.0695 |

The dynamic Leidenfrost temperature ($T_{DL}$) of ferrofluid droplets in the absence of magnetic field were determined by conducting benchmark experiments at different Weber numbers, $We_o = \rho V_o^2 D_o / \sigma$ (based on free/zero-field droplet impact velocity, $V_o = \sqrt{2gh_o}$ ) and nanoparticle concentrations, where, $\rho$, $D_o$, $\sigma$, and $h_o$ are the ferrofluid density, pre-impact droplet diameter, surface tension, and droplet release height, respectively. The $T_{DL}$ was considered as the minimum substrate temperature at which droplet rebound was observed for the first time with minimal spraying [10-13] and the obtained values are given in Table 2. To explore magnetic field effects, the magnetic Bond number, $Bo_m = \chi B_o^2 D_o / \mu_o \sigma$ (ratio of magnetic force to surface tension force), was varied by changing the applied magnetic field strength and nanoparticle concentration, where, $B_o$, $\mu_o$, and $\chi$ are the imposed magnetic flux density, magnetic permeability of vacuum and magnetic susceptibility of ferrofluids. The values of $\chi$ were adopted from the saturation susceptibilities reported by Rigoni et al. [54]. The lower limit of $h_o$= 3 cm was chosen to minimize the influence of magnetic field on droplet dispensing, while the upper limit of $h_o$ = 7 cm was selected to avoid droplet fragmentation immediately upon impact. The magnetic field strength was varied from 0 to 200 mT. Beyond 200 mT, the magnetic field became sufficiently strong to induce significant lateral droplet motion, resulting in oblique impact on the substrate. In line with the previous studies [10-13], the onset condition for the dynamic Leidenfrost state in the presence of magnetic fields was identified as the minimum magnetic field strength (and corresponding $Bo_m$) at which droplets first rebounded from the substrate with minimal spraying.

**Table 2:** Dynamic LFT of ferrofluid droplets in the absence of magnetic field.

| Release height, $h_o$ (cm) | Dynamic Leidenfrost temperature, $T_{DL}$ (°C) | |
|---|---|---|
| | 5% w/w | 7.5% w/w |
| 3 | 190 | 180 |
| 5 | 180 | 175 |
| 7 | 170 | 170 |

## 3. Results and discussion

The discussion is organized into six subsections. We begin by investigating magnetic field effects on droplet impact at $T_s$ = 160 °C, a temperature below the minimum zero-field dynamic Leidenfrost temperature (see Table 1). We identify the criteria governing different outcomes across a range of impact conditions and nanoparticle concentration, and report a novel phenomenon, namely the magneto-Leidenfrost effect (MLFE). We then investigate the magnetic-field-induced modifications to the Leidenfrost droplets at elevated $T_s$ = 180 °C ($\geq T_{DL}$). Subsequently, at $T_s$ = 225 °C, well above $T_{DL}$, we examine droplet boiling behaviour in the fragmenting Leidenfrost regime. Finally, a theoretical model is presented to explain the experimental observations and provide mechanistic insights into the role of magnetic forces in governing the thermo-fluid-dynamics.

### *3.1. Effect of magnetic field on the onset of the dynamic Leidenfrost effect*

Fig. 2. Illustrates the temporal evolution of ferrofluid droplets impacting the hot substrate at $T_s$ = 160 °C, for different $Bo_m$ and $We_o$ at a nanoparticle concentration of 5% w/w. The effect of magnetic field is shown in Figs. 2(a), (b) and (c) for $We_o$ ~22, 37, and 56, respectively. Columns (i)-(v) illustrate different stages (with respective time-stamps) of the impact dynamics. Since the substrate temperature is below the zero-field $T_{DL}$ values, the droplets in the absence of magnetic fields ( $Bo_m = 0$) remain attached to the substrate at the end of recoil and continue to vaporize, as depicted by first rows from the top in Fig. 2. However, in the presence of sufficiently strong magnetic fields, the droplets start to rebound from the substrate at the end of recoil phase, as shown in Figs. 2(b) and (c). For $We_o \approx 37$, as illustrated in Fig. 2(b), when the $Bo_m$ is increased to 36.3 and beyond (by increasing magnetic field strength) the ferrofluid droplets show rebounding behaviour. This is direct evidence for magnetic field induced onset of the Leidenfrost regime, which we name as the MLFE.

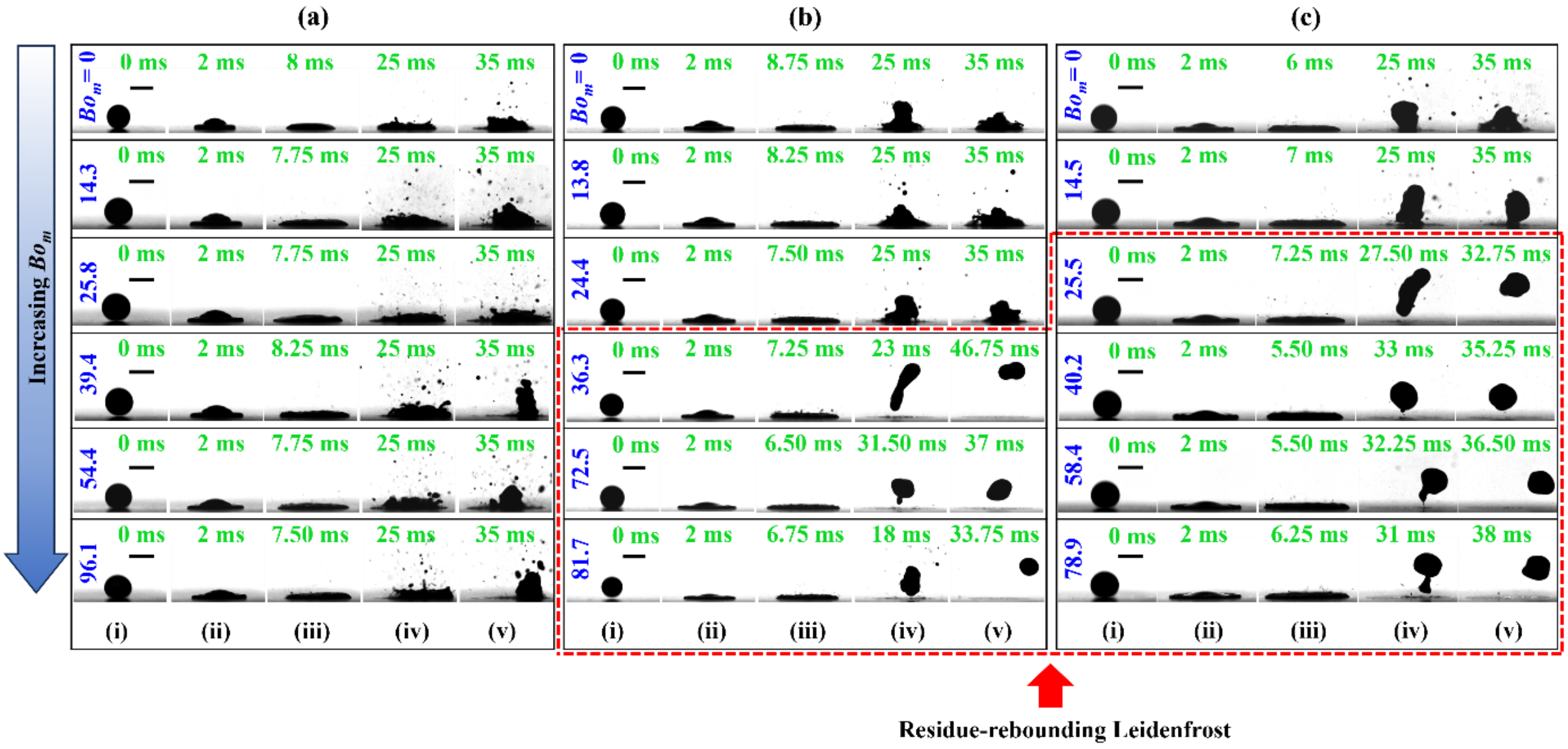

**Fig. 2.** Time-lapse depiction of the effect of magnetic field on the impact dynamics and boiling behaviour of ferrofluid droplets of 5% w/w nanoparticle concentration at substrate temperature $T_s = 160$ °C. (a) At $We_o \approx 22$, droplets remain attached to the substrate. (b) At $We_o \approx 37$, the onset of the dynamic Leidenfrost effect occurs at $Bo_m \geq 36.3$, resulting in droplet rebound. (c) A further increase to $We_o \approx 56$ lowers the threshold $Bo_m$ required for the onset of rebounding, exhibiting the MLFE. The scale bars represent 2.5 mm. Column (iii) depicts droplets at maximum state stage, while for rebounding cases, column (iv) shows the instant of droplet lift-off from the substrate. The region enclosed by the dashed lines illustrates the MLFE-driven rebound regime.

In principle, for the Leidenfrost effect to prevail, the vapour pressure underneath the droplet should be sufficiently high to eliminate the droplet-substrate contact. For gently deposited droplets ($We_o$ ≈ 0), the scale of vapour pressure is obtained using a lubrication model assuming the outward vapour flow in the thin vapour layer as Poiseuille flow [14,50], and is found to be proportional to the square of droplet contact diameter. Also, an increase in contact diameter means area available for heat transfer is more and availability of more nucleation sites, resulting in more production of vapour, favouring the LFE. In our study, the consideration of dynamic impact effects becomes essential. On one hand, the higher inertia of impinging droplets results in larger spread diameter, which eventually favours the LFE by increasing the vapour pressure [12,13,34-37]. On the contrary, however, for droplets with high impact inertia to achieve LFE, the vapour pressure must be large enough to balance the dynamic pressure arising due to impact momentum, and may require higher substrate temperatures [23,32,33] to avoid splashing/fragmentation immediately post-impact. An experimental study by Mitchell et al. [55] revealed that the impact force is significant only for a very short time-interval during impact. This indicates that a more favourable condition for the Leidenfrost effect manifests during later stages of spreading, when impact momentum has reduced to a negligibly small value. Therefore, for moderate impact conditions, as considered in our work ($We_o$ ~ 20-56), where droplets always survive the impact, the increase in $We_o$ results in lower $T_{DL}$. This also explains the observed $T_{DL}$ values from our benchmark experiments (see Table 2), and is consistent with literature [12,13].

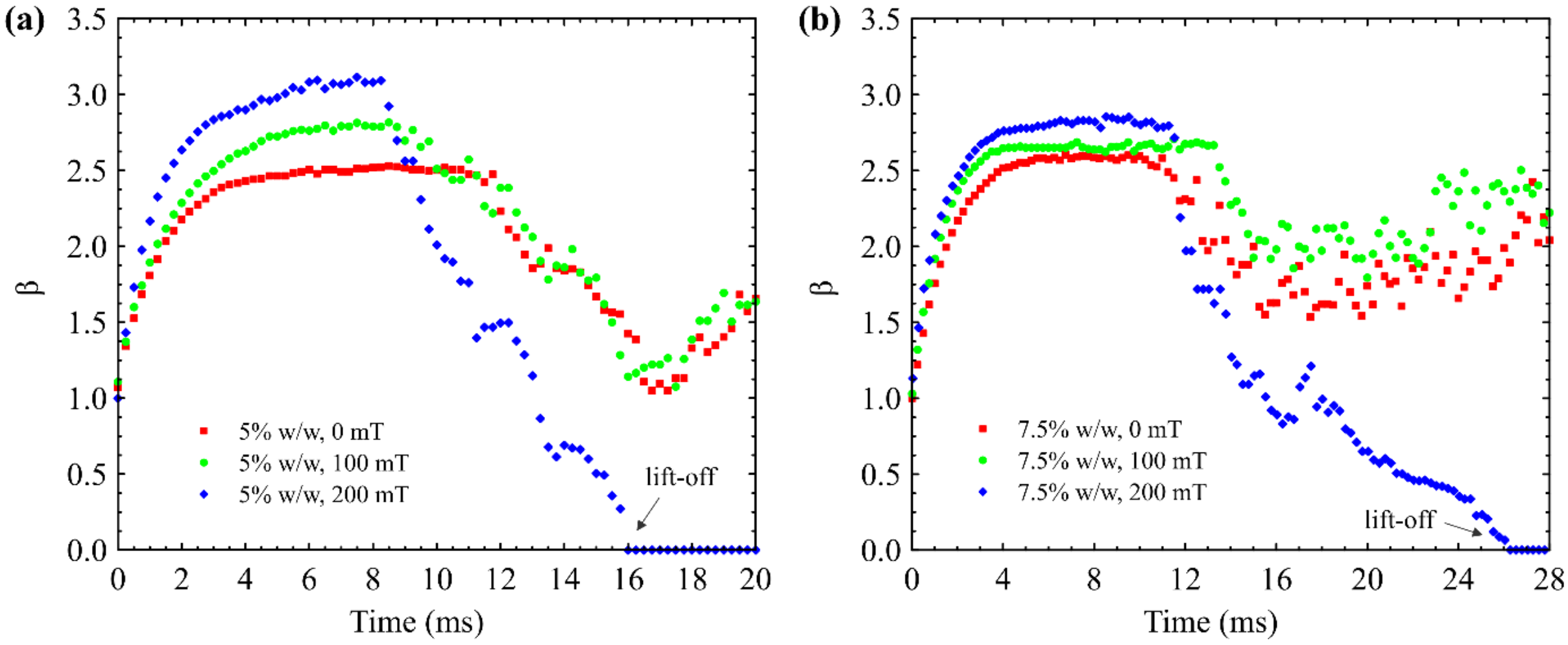

**Fig. 3.** Temporal evolution of spread factor, $\beta = D / D_o$, at different magnetic field strengths for ferrofluid droplets with nanoparticle concentrations of (a) 5% w/w and (b) 7.5% w/w, at $We_o$≈ 37 and $T_s$ = 160 °C. A substantial enhancement in droplet spreading is observed with increasing magnetic field strength. The cases corresponding to $B_o$ = 200 mT exhibit droplet lift-off at the end of retraction phase, indicating the MLFE.

To quantify the spreading diameter, the spread factor $\beta = D / D_o$ is defined, which is the ratio of instantaneous droplet diameter to the pre-impact diameter. Figs. 3(a) and (b) show the spreading behaviour of ferrofluid droplets at different magnetic field strengths for $h_o$ = 5 cm ($We_o \approx$ 37) at $T_s$ = 160 °C, for 5% w/w and 7.5% w/w concentrations, respectively. It shows that an increase in magnetic field strength results in enhanced spreading, leading to larger maximum spread diameter. This is attributed to the magnetic forces arising due to the interaction of ferrofluid droplets with externally imposed field, promoting the spreading by first increasing the impact velocity by pulling it downwards, and then horizontally stretching it away from the point of impact; a detailed discussion on this is given in the Section 3.6.

Enhanced spreading due to magnetic forces results in increase in heat transfer from the substrate, due to increase of the contact area, and the number of nucleation sites available for vapour inception. Consequently, the vapour pressure underneath increases and promotes the formation of a stable vapour layer. Therefore, beyond a threshold magnetic field, the formation of vapour layer allows droplets to recoil back smoothly and eventually rebound from the substrate. The formation of the stable layer is further proven by the fact that at higher fields, despite the spread factor being large, the duration the droplet spends in this state diminishes largely. In Figs. 3(a) and (b), the $B_o$ = 200 mT cases illustrate impact-outcomes in which droplets undergo rebound/lift-off, whereas at $B_o$ = 0 and 100 mT they remain attached to the substrate. Since, here, the presence of magnetic field explicitly triggers the onset of LFE-driven droplet rebounding at a substrate temperature below its usual $T_{DL}$ (zero-field Leidenfrost temperature), we name this phenomenon as MLFE. For a higher $We_o$, the droplet with already a large spread diameter due to high impact inertia requires a relatively smaller magnetic field (smaller $Bo_m$) for the onset of droplet rebound stage. Consequently, droplets begin to show rebound behaviour at a relatively lower threshold value of $Bo_m$ = 25.5 for larger $We_o$ = 56 (larger release height), as shown in Fig. 2(c). It is noteworthy that for 5% w/w droplet, we did not observe any rebounding behaviour for the lowest considered value of $We_o \approx$ 22. This could be because of $We_o$ being too low, resulting in very small spreading diameter. Therefore, much larger magnetic forces would have been needed for the droplets to rebound, which however disrupts the droplet away from its location on the substrate and generates artefacts.

### *3.2. Effect of increase in magnetic nanoparticle concentration*

As a colloidal droplet spreads over a hot substrate, nanoparticles near the liquid-substrate interface deposit onto the surface, eventually forming a thin layer of nanoparticle residue. This can be observed in Figs. 2(b) and (c) by carefully looking at the substrate top surface in column (v) of all the rebounding cases. This surface-attached nanoparticle residue significantly modifies the surface microstructure, and strongly affects the droplet recoil behaviour. An increase in nanoparticle concentration produces denser nanoparticle residues, and increases the number of nucleation sites, thereby further promoting bubble incipience. Moreover, a decrease in average gap between residue particles enhances the chances of incipient bubble coalescence, and subsequent formation of stable vapour layer. Also, denser residue/surface-protrusions weakens the flow of escaping vapour, thereby increasing vapour pressure underneath, a similar effect as observed by Kim et al. [40], of decrease in Leidenfrost temperature at higher micropillar density on micro-structured surfaces. Consequently, in contrast to the 5% w/w droplets, rebounding is observed for the 7.5% w/w droplets even at the lowest $We_o \approx 20$, as illustrated in Fig. 4(a). Furthermore, consistent with the trend observed for the 5% w/w droplets, the critical $Bo_m$ for the onset of the MLFE decreases with increasing $We_o$ for the 7.5% w/w droplets as well, (Figs. 4(a) and (b)).

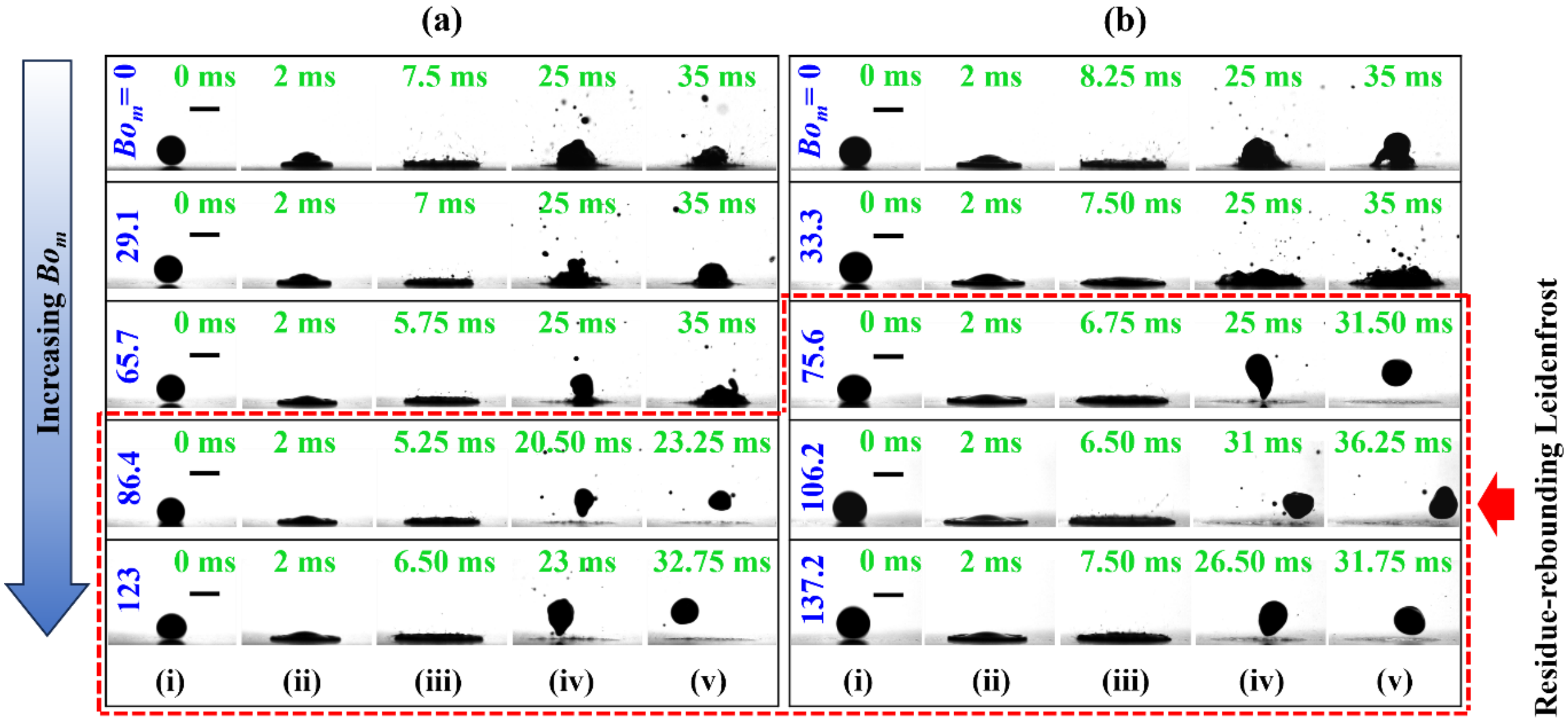

**Fig. 4.** Effect of magnetic field on the impact dynamics and boiling behaviour of ferrofluid droplets of 7.5% w/w at $T_s = 160$ °C. (a) At $We_o \approx 20$, rebound is observed for $Bo_m \geq 86.4$, unlike in the 5% w/w case. (b) Increasing $We_o$ to 40 lowers the threshold $Bo_m$ required for the onset of rebound. The scale bars represent 2.5 mm. Column (iii) depicts droplets at maximum spread stage, while for rebounding cases, column (iv) shows the instant of droplet lift-off. The region enclosed by the dashed lines illustrates the MLFE regime.

The 7.5% w/w droplets exhibited a greater tendency to fragment after impact. Therefore, only two droplet release heights $h_o = 3$ and 5 cm ($We_o \approx 20$ and 40) were considered. Finally, to delineate the MLFE, the critical $Bo_m$ was used as the demarcating parameter separating the no-lift-off and rebounding outcomes, as indicated by the red dashed boundaries in Figs. 2 and 4. Following the terminologies used in the literature [13], this Leidenfrost boiling mode is classified as the residue-rebounding Leidenfrost mode, which refers to the dynamic LFE of impinging colloidal droplets.

### *3.3. Effect of magnetic field on different stages of Leidenfrost dynamics*

Fig. 5(a) shows the variation of experimental droplet impact velocity $V_i$ with magnetic field strength for different release heights, due to the field pulling on the falling magnetic-fluid. Here, the impact velocity increases with magnetic field strength for a fixed release height. Further, for a fixed magnetic field strength and release height, this increase is larger for 7.5% w/w droplets in comparison to 5% w/w cases. The higher population of magnetic nanoparticles lead to higher magnetization, therefore stronger magnetic force induce higher downward pull. Later we show that the non-uniformity of the magnetic field in the study domain gives rise to magnetic forces, acting in the vertically downward direction, accelerating the droplet descent until impact, as well as in the radially outward direction in the droplet spreading plane, influencing its post-impact dynamics. A detailed discussion on this is provided in Section 3.6. Hence the net increased spreading (see Fig. 3) results from the combined effect of increased impact velocity due to the vertical magnetic forces, and the outward radial stretching caused by horizontal magnetic forces.

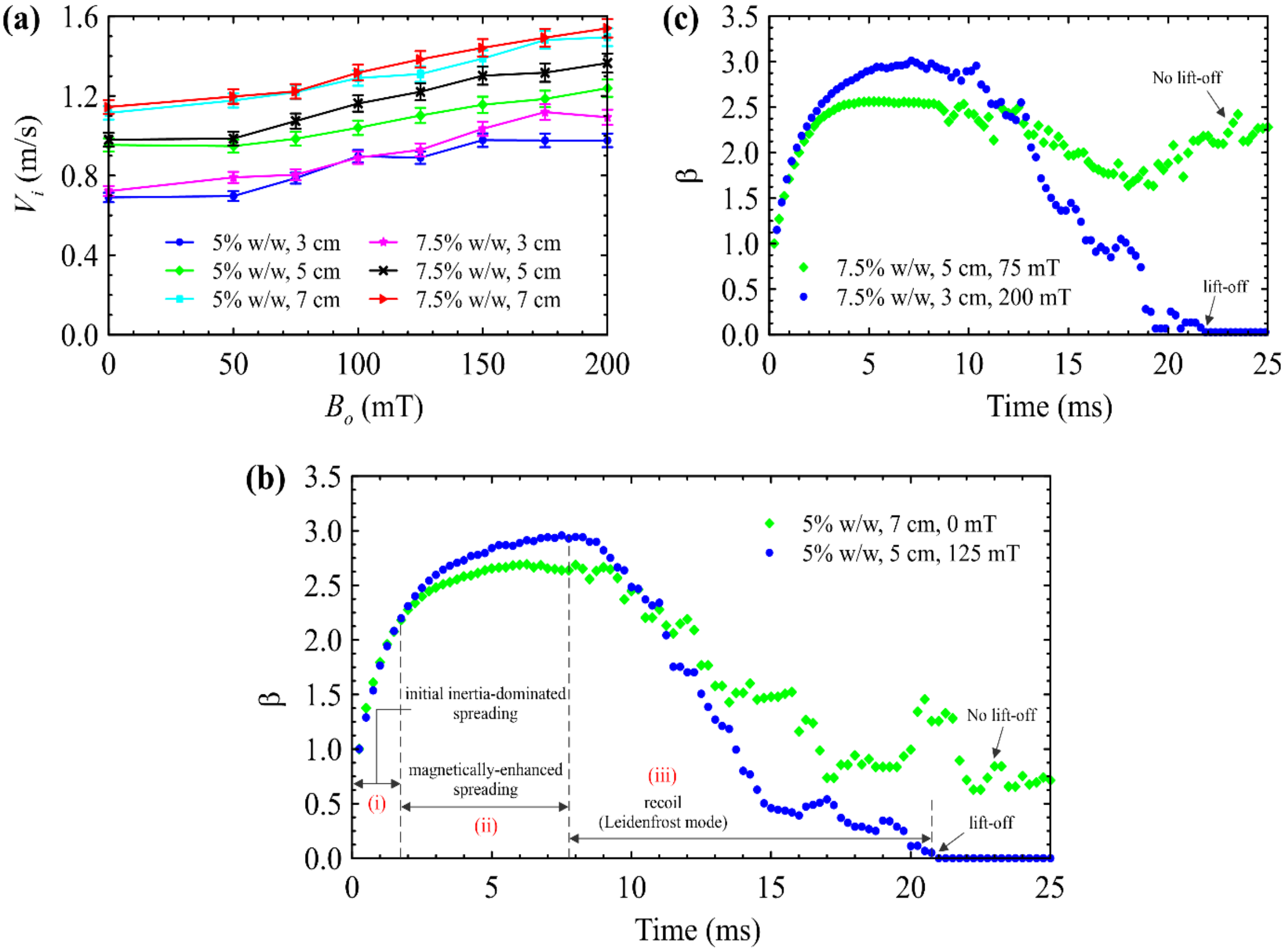

**Fig. 5.** Effect of magnetic field on the droplet impact velocity, $V_i$, and subsequent impact dynamics: (a) Variation of $V_i$ with $B_o$ for different release heights and nanoparticle concentrations, (b, c) evolution of $\beta$ during different stages of spreading, retraction, and subsequent lift-off/no-lift-off for 5% w/w and 7.5% w/w droplets, respectively.

We consider two scenarios, (see Figs. 5(b) and (c)). First, 5% w/w droplets with identical impact conditions (equal impact velocities), one at zero-field and the other with imposed-field conditions are compared. Fig. 5(b) illustrates the post-impact spreading behaviours of ferrofluid droplets of equal impact velocities, with and without magnetic fields for $T_s$ = 160

°C. The 5% w/w, 7 cm, 0 mT case droplet remains attached to the substrate, whereas 5% w/w, 5 cm, 125 mT case shows MLFE rebounding/lift-off. In the absence of field, after being released from $h_o$ = 7 cm, the droplet descends freely (assuming negligible air drag) and attains an impact velocity of $V_o = \sqrt{2gh_o}$ . At lower release height of 5 cm, a magnetic field of 125 mT strength is imposed to achieve the same impact velocity.

With impact kinetic energies being identical, after impinging on the substrate, the droplets in both cases spread in a similar fashion in the early inertia-dominated spreading regime-(i) ($t$ < 1.8 ms). Thereafter, the horizontal magnetic forces acting radially outward enhance droplet spreading, as shown by magnetically-enhanced spreading regime-(ii) (see Fig. 5 (b)). Finally, the capillary forces trying to minimize the droplet surface area overcome the magnetic forces and the recoil phase begin (see recoil regime-(iii)). For the non-zero field case, due to enhanced spreading, the vapour layer sets-in (as explained earlier) and MLFE is achieved, and the droplet recoils smoothly over a stable vapour layer and undergoes lift-off. For the zero-field case, low spreading, and thus the absence of vapour layer formation prohibits any lift-off, as the retraction phase leads to viscous dissipation at the substrate. This clearly indicates that despite identical impact velocities, the application of a magnetic field does induce LFE-driven droplet rebound. The observations are cemented by similar MLFE dynamics of the 7.5% w/w droplets (see fig. 5(c)).

### *3.4. Residence time in the rebounding Leidenfrost regime*

To quantify the rebound dynamics, we examine the droplet residence time, defined as the time interval between the first droplet-substrate contact and first complete rebound from the substrate. The residence time is an important parameter for heat transfer considerations, as it represents the duration available for effective heat transfer between the substrate and fluid. In applications involving droplet actuation and transport [56,57], reduced residence time is desirable for rapid droplet motion while minimizing liquid mass loss due to heat transfer. A substrate temperature of 180 °C ($\geq T_{DL}$) was chosen such that it is sufficiently superheated to always induce droplet rebound, even in the absence of a magnetic field for both 5% w/w and 7.5% w/w droplets. Fig. 6 shows the temporal evolution of the post-impact dynamics of droplets in the LFE regime for $We_o \approx 38$. The droplet rebound instant is depicted in column (iv), with the time stamp indicating the residence time ($t_{res}$).

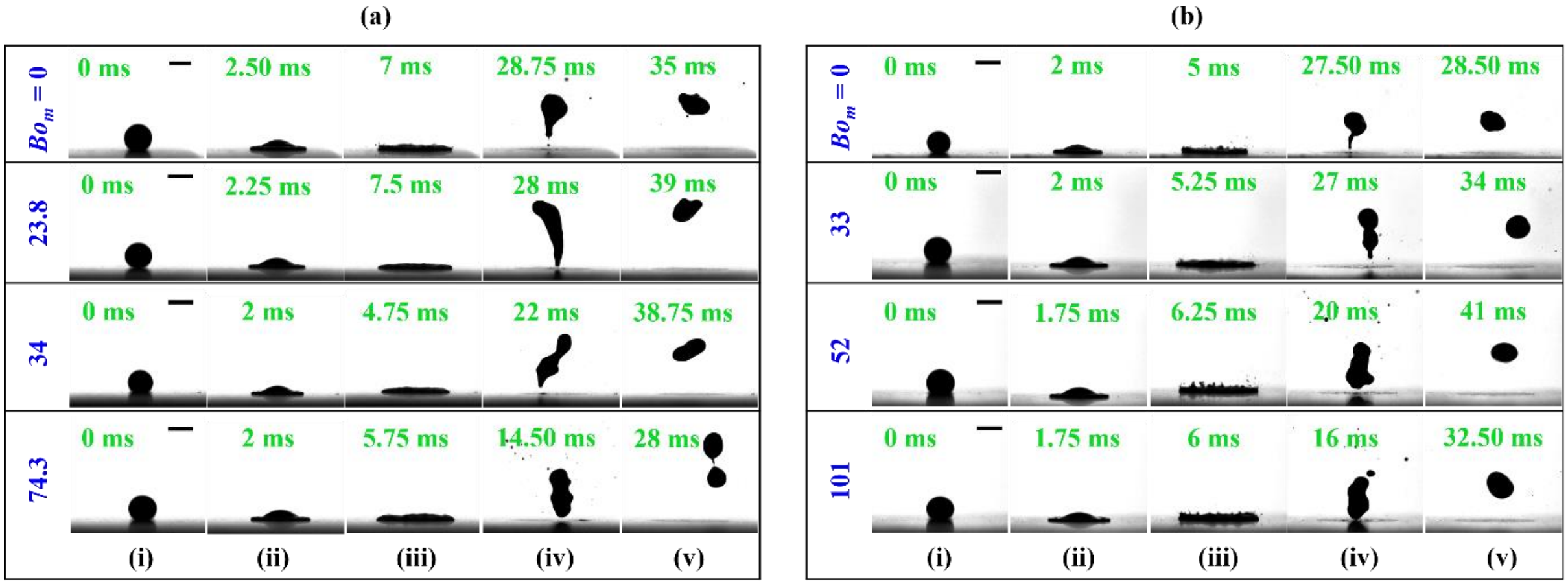

**Fig. 6.** (a, b) Evolution of MLFE droplets with concentrations of 5% w/w and 7.5% w/w, respectively, for varying $Bo_m$, at $T_s$ = 180 °C and $We_o \approx 38$. The scale bars represent 2.5 mm. The time stamp shown in column (iv) depicts droplet residence time.

At $Bo_m = 0$, as depicted by the top rows, the droplets rebound from the substrate in ~28.75 ms and ~27.50 ms for 5% w/w and 7.5% w/w cases, respectively. However, the $t_{res}$ decreases to as low as ~14.50 ms and ~16 ms at $Bo_m$ = 74.3 and 101 for the 5% w/w and 7.5% w/w cases, respectively. This is attributed to the existence of more pronounced and stable Leidenfrost state in the presence of magnetic fields, the MLFE, which further reduces frictional effects producing prompt rebound. This is evident from Figs. 6(a) and (b), as the time gap between the maximum spread stage (column (iii)) and the lift-off stage (column (iv)) decreases with $Bo_m$. Previous studies [58-61] have shown that the residence time of droplets impinging on superhydrophobic and/or hot surfaces in Leidenfrost regime is largely independent of the impact velocity, and can be approximated as the period of a freely oscillating droplet ((Rayleigh time-period)) [62] as,

$$\tau_R = \frac{\pi}{4}\sqrt{\frac{\rho D_o^3}{\sigma}} \tag{1}$$

This timescale, though without a pre-factor $\pi/4$, can also be obtained by balancing the inertia with capillarity in the low-viscosity limit. Herein, we normalise the residence time by the $\tau_R$ to obtain a dimensionless residence time $\tau_{res}$,

$$\tau_{res} = \frac{t_{res}}{\tau_R} \tag{2}$$

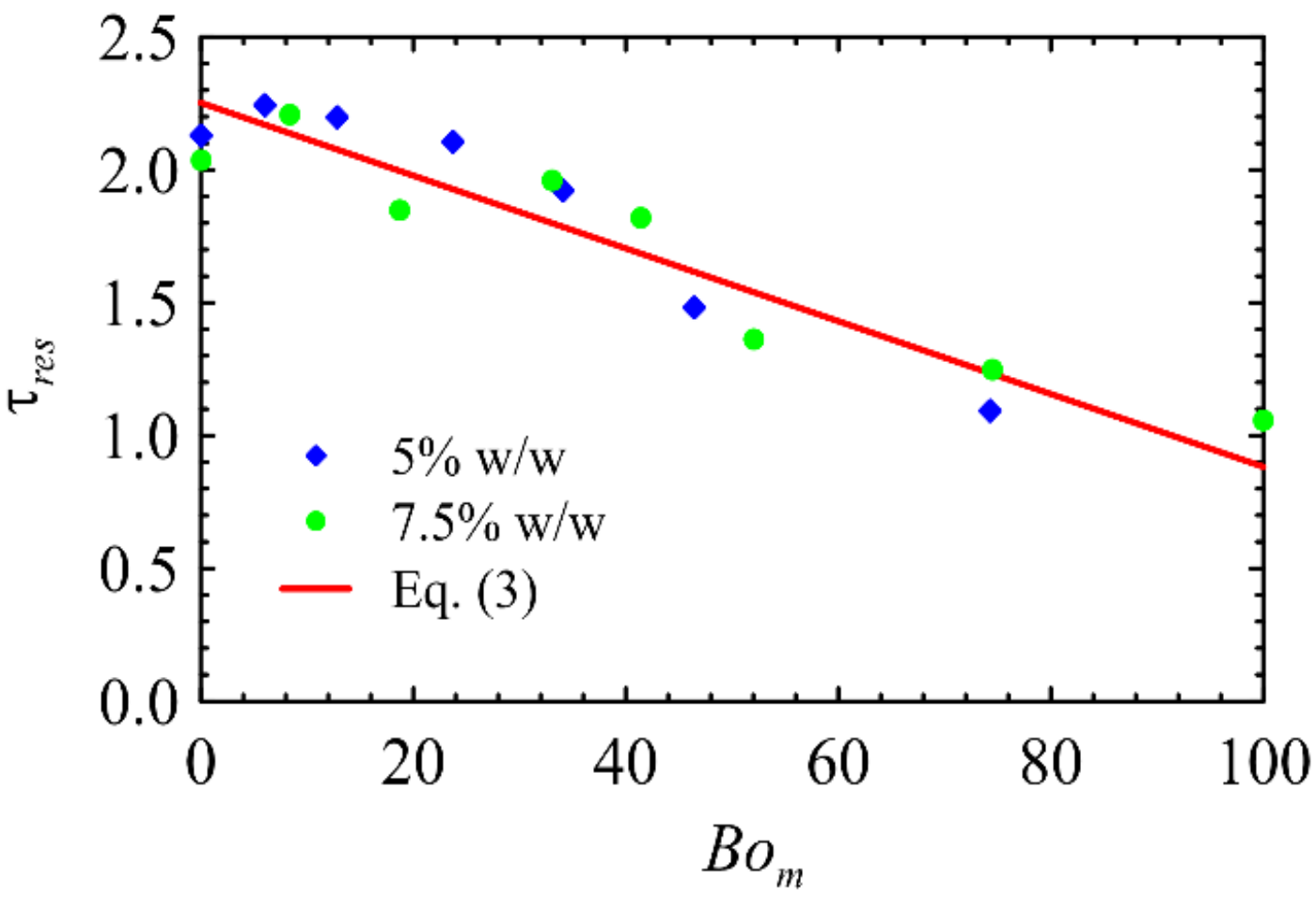

**Fig. 7.** Variation of the with $Bo_m$ in the Leidenfrost regime at $T_s$ = 180 °C and $We_o \approx 38$. The solid line represents the regression Eq. (3), indicating a linear dependence of the residence time on $Bo_m$.

As shown in Fig. 7, at low $Bo_m$, the $\tau_{res}$ remains significantly >1, and decreases monotonically with increase in $Bo_m$. For higher $Bo_m$, it decreases substantially and approaches 1 ($\tau_{res} \approx \tau_R$). We note that the dimensionless residence times at high $Bo_m$ are close to those reported for droplet rebound from superhydrophobic or sufficiently hot surfaces ($T_s >> T_{DL}$) [57,61,63]. Interestingly, the $\tau_{res}$ for both the 5% w/w and 7.5% w/w cases approximately collapse onto a single straight line (see Fig. 7), obtained through least-squares fit with a mean relative error of ~7%, as

$$\tau_{res} = 2.2526 - 0.0137\, Bo_m \quad (3)$$

This is an important outcome, as it tells that magnetic field not only provide better control over droplet rebound time but also offers flexibility in selecting different combinations of ferrofluid composition and magnetic field strength to meet application requirements. Such flexibility could be particularly beneficial in applications constrained by the permissible magnetic field strength, or the allowable concentration of foreign components (e.g., magnetic nanoparticles) within the fluid.

### *3.5. Effect on boiling behaviour in high temperature, fragmenting Leidenfrost regime*

At $T_s$ = 225 °C (well above $T_{DL}$), the droplet undergoes fragmenting Leidenfrost boiling (see Fig. 8(a)). Due to vigorous boiling at large superheat, the droplet fragments into multiple daughter droplets, whose size and distribution depend on the magnetic field strength. For a relatively low $Bo_m = 24.3$, the droplet spreads over a thick vapour layer formed by rapid evaporation upon impact, while vapour escapes radially outward. During spreading, interfacial instability at the rim produces protruding liquid filaments/fingers that grow under inertia and are further stretched outward by magnetic forces (see outward pointing red arrows in $Bo_m = 24.3$, column (iv)). This leads to the ejection of secondary droplets from the rim (see green arrows in column (v)) while the remaining lamella retracts see inward pointing green arrows in column (iv)) and rebounds from the substrate.

As $Bo_m$ increases to 81.7, the field-aided increased impact velocity causes rapid spreading, following a brief contact with the substrate, leaving insufficient time for radial vapour escape. Consequently, vapour becomes trapped beneath the droplet during the early stages of spreading ($t = 1.75$ ms; depicted by green arrow in $Bo_m = 81.7$, column (ii)). As the lamella stretches and thins, the trapped vapour escapes in pockets, and destabilizes and ruptures the lamella at multiple locations. As shown in columns (iii)-(iv) for $Bo_m = 81.7$, multiple nucleated holes appear and grow within the lamella, producing a filamentous structure where the outer rim remains connected to the central liquid region through spoke-like ridges. Near maximum spread state, both the rim and the connecting ridges rupture into multiple big daughter droplets and smaller bridge droplets. Also, at $Bo_m = 81.7$, the lamella rupture and subsequent hole growth cause early retraction of the central liquid, thereby limiting the growth of the rim as well as its protruding filaments. Thus, the ejection of secondary droplets is suppressed.

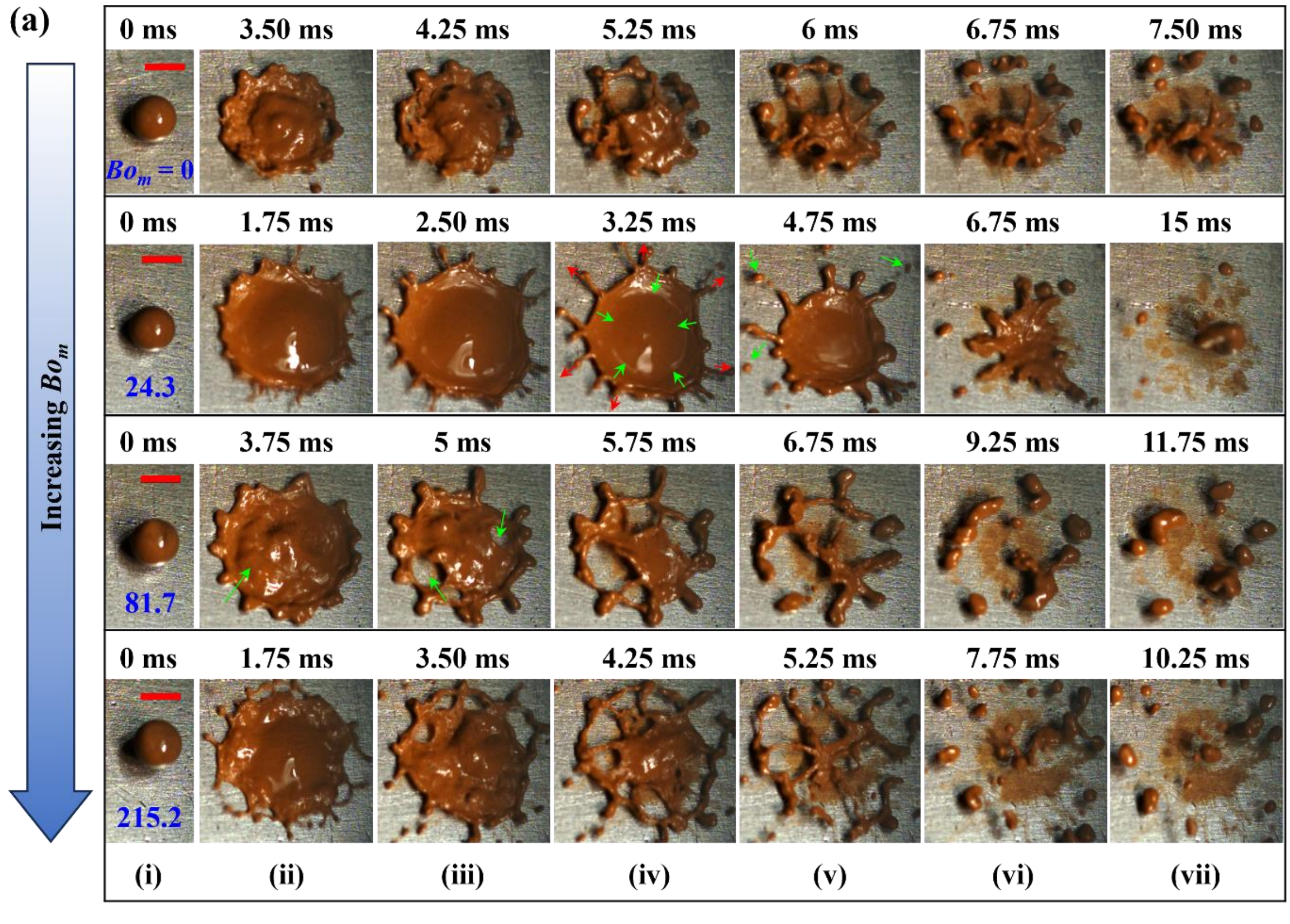

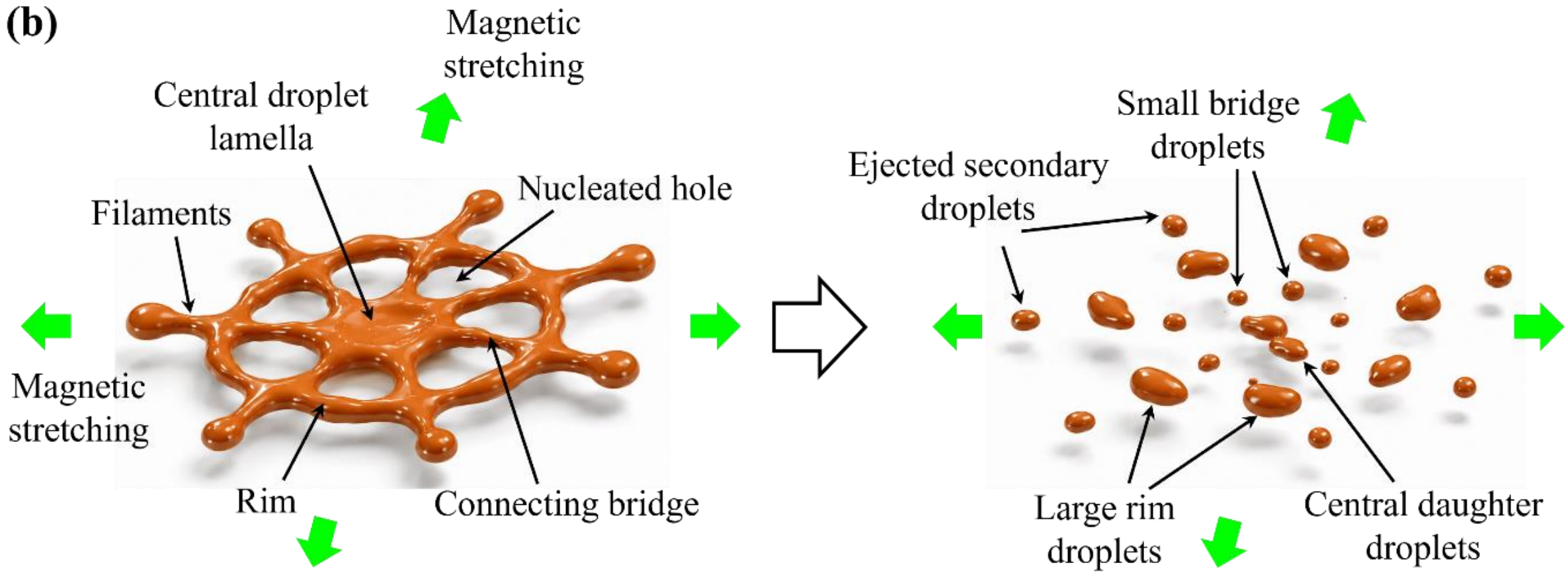

**Fig. 8.** (a) Top-view images of temporal evolution of ferrofluid droplets during fragmenting Leidenfrost boiling, with increasing $Bo_m$ . For $Bo_m = 24.3$, the outward red arrows in column (iv) show the growth of filaments from outer rim, while the inwards green arrows show the retraction of central lamella. The green arrows in column (v) highlight rim-ejected secondary droplets. For $Bo_m = 81.7$, the arrows in columns (ii) and (iii) show locations of trapped vapour beneath the lamella and the subsequent nucleation of holes within the lamella, respectively. (b) Illustrative image (AI generated from experimental image) showing the different components of filamentous structure and the various types of daughter droplets generated during fragmenting Leidenfrost boiling at $T_s$ = 225 °C.

At high $Bo_m$ = 215.2, a stronger horizontal magnetic pull in combination with higher impact inertia led to reappearance of secondary droplet ejection from rim, despite more violent and early lamella rupture. Due to higher stretching, the droplet fragments at multiple locations, including the protruding filaments, rim, connecting ridges, and central lamella. As a result, numerous daughter droplets of varying sizes and types are generated, including filament-ejected secondary droplets, fragmented rim-droplets, central daughter droplets, and small bridge droplets (see Fig. 8(b)). In the absence of magnetic fields ($Bo_m$ = 0), an intermediate behaviour is observed. Here, part of the trapped vapour ruptures the lamella locally, while the remaining vapour escapes radially outward, resulting in droplet fragmentation without noticeable protruding filaments from the rim.

### *3.6. Magnetic force scaling and prediction of maximum spread factor*

Fig. 9(a) illustrates two stages of a ferrofluid droplet released from a height $h_o$ and moving through a magnetic field environment: (i) its downward descent prior to impact, at a vertical distance $h$ above the substrate located at the origin, $O\ (r=0,\ \theta=0,\ z=0)$, and (ii) during post-impact spreading. A cylindrical coordinate system $(r,\ \theta,\ z)$ is adopted to describe the magnetic field distribution in the study domain, where $B_r$, $B_\theta$ and $B_z$ denote its radial, azimuthal, and axial components, respectively. Owing to the geometrical symmetry about the pole axis ($r$ = 0), $B_\theta$ and its gradients are zero. Away from the pole axis, $B_r$ becomes finite, due to the radial outward bending of field lines. The field lines bend back and become horizontal at $z$ = 0, resulting in $B_r$ = 0 at this location. Further, along the pole axis, the field lines remain parallel to the $z$-direction, resulting in $B_r$ = 0 for $r$ = 0. Thus, at the origin $B_z = B_o$, where $B_o$ denotes the net magnetic field strength at the origin.

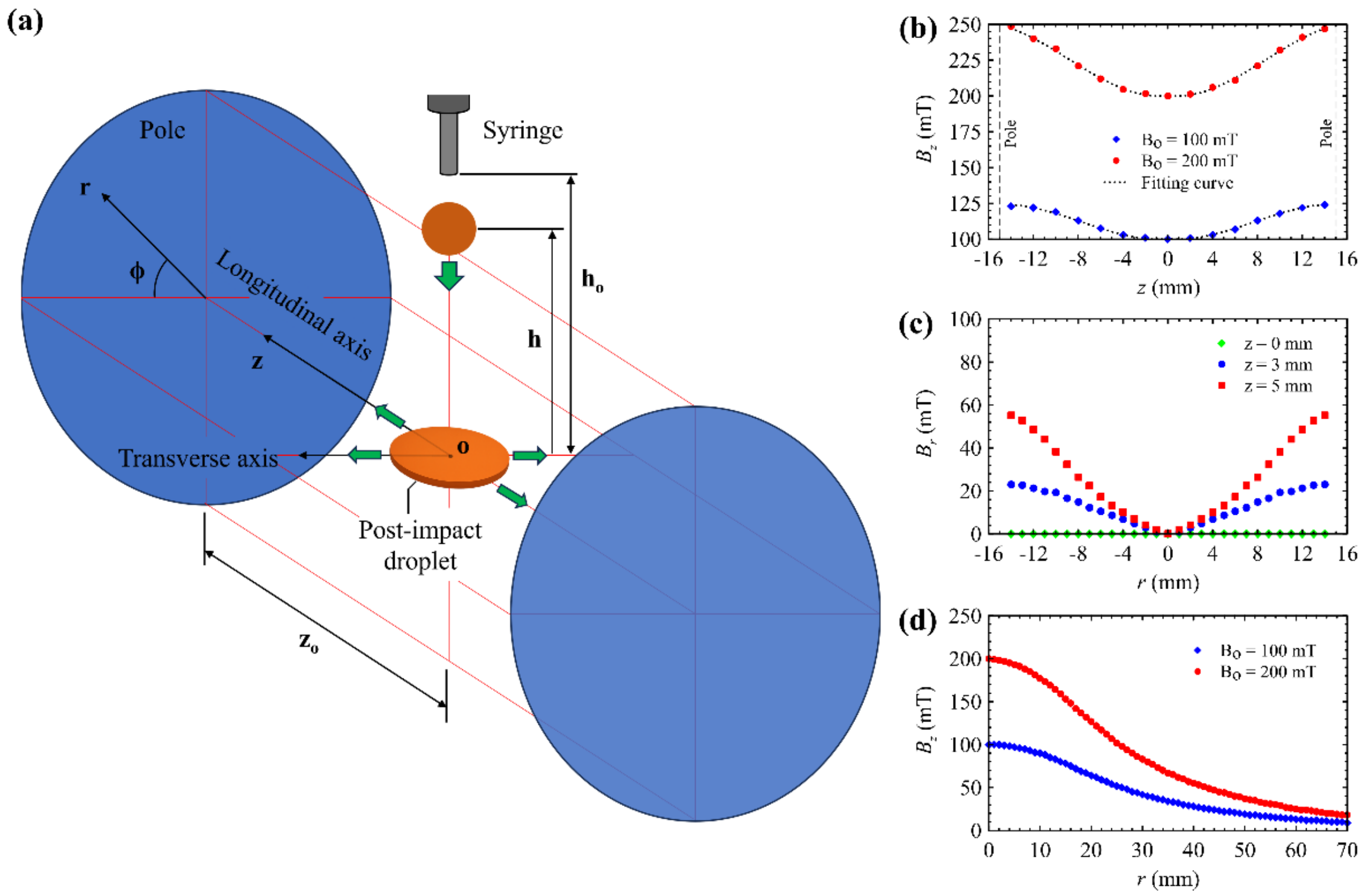

**Fig. 9.** (a) Schematic of the experimental domain used for theoretical modelling, (b) variation of $B_z$ along the pole axis. (c) Variation of $B_r$ along the radial direction in the vicinity of the impact location. (d) Variation of $B_z$ along the radial or vertical *h*-direction away from the substrate.

Figs. 9(b) and (c) show the magnetic field distribution in the vicinity of the origin, where droplet impact and spreading occur. Fig. 9(b) shows the variation of measured values of $B_z$ along the pole axis for $B_o$ = 100 and 200 mT, showing that $B_z$ decreases away from the poles and becomes equal to $B_o$ at $z = 0$. Fig. 9(c) shows the variation of $B_r$ along the radial direction at $z = 0$, 3, and 5 mm. It reveals that the $B_r$ increases sharply with r and becomes larger for higher *z*-coordinates due to bending of the field lines. Fig. 9(d) shows the variation of $B_z$ with *r* along the vertical h-direction (parallel to the droplet motion prior to impact), at $z = 0$ for $B_o$ = 100 and 200 mT. The axial magnetic field decreases monotonically with increasing distance from the substrate. Notably, after an initial gradual decrease (for small *r*), it decays rapidly with height and asymptotically approaches zero as the magnetic field lines become increasingly sparse.

When the ferrofluid droplet comes within the influence of magnetic field, it gets magnetized. The magnetic energy per unit volume is [46] as:

$$E_m = -\frac{\chi B^2}{2\mu_o} \quad (4)$$

Where $B$, $\mu_o$, and $\chi$ are applied magnetic field, permeability of free space and magnetic susceptibility of the ferrofluid, respectively. The derivative of the magnetic energy yields magnetic force per unit volume [52] as:

$$F = \frac{\chi}{2\mu_o}\nabla(B^2) \quad (5a)$$

$$F_r = \frac{\chi}{\mu_o}\left[B_r\frac{\partial B_r}{\partial r} + B_z\frac{\partial B_z}{\partial r}\right], \qquad F_z = \frac{\chi}{\mu_o}\left[B_r\frac{\partial B_r}{\partial z} + B_z\frac{\partial B_z}{\partial z}\right] \quad (5b)$$

Where, $F_r$ and $F_z$ are conservative magnetic forces per unit volume along *r* and *z* direction.

After being released from a height $h_o$, the droplet descends vertically downwards towards substrate at origin ($r = 0, z = 0$). Magnetic field lines being mostly horizontal near $z \approx 0$, implies, $B_r \approx 0$ and $\partial B_r/\partial r \approx 0$. Therefore, the vertical magnetic force acting on the ferrofluid droplet is expressed as,

$$F_v = F_r \approx \frac{\chi}{\mu_o}\left[B_z\frac{\partial B_z}{\partial r}\right] \quad (6)$$

Since, $\partial B_z/\partial r$ is negative, as evident from Fig. 9(d), the droplet experiences a vertically downward force pulling it towards the substrate. The nature of the side-wise force $F_z$ promotes lateral stretching along *z*-direction, and deforms the droplet into an oblate shape. However, within the range of magnetic fields considered herein, no considerable deviation from spherical droplet shape was noted.

Assuming, the variation of magnetic field along the vertical *h*-direction to be linear, of form $B_z = B_o(1 - h/h_o)$, the work done by vertical magnetic forces is expressed as,

$$W_{m_d} = \int_{h_o}^{0} F_v\, dh = -\frac{\chi}{\mu_o}\int_{h_o}^{0}\frac{B_o^{\,2}}{h_o}\left(1-\frac{h}{h_o}\right)dh = \frac{\chi B_o^{\,2}}{2\mu_o} \quad (7)$$

This magnetic work aids to further accelerate the droplet towards substrate, thereby increasing its impact kinetic energy, given by:

$$KE_1 = \left(\rho g h_o + \frac{\chi B_o^2}{2\mu_o}\right) \times \frac{1}{6}\pi D_o^3 \tag{8}$$

And, the surface energy of the droplet just before impact is given by,

$$SE_1 = \sigma \pi D_o^2 \times \frac{1}{6}\pi D_o^3 \tag{9}$$

In the presence of magnetic fields, the spreading of ferrofluid droplets is governed by the interplay of capillary, inertial, viscous, and magnetic forces, with the droplet assuming a pancake-like shape at the maximum spread state. The axial symmetry results in an identical distribution of the magnetic field in both the vertical (as illustrated in Figs. 9(b), (c), and (d)) and horizontal planes, corresponding to the droplet descent and spreading phases, respectively. During spreading, the net magnetic forces act within the droplet spreading plane along two orthogonal directions: the longitudinal direction (parallel to the *z*-axis) and the transverse direction (normal to the *z*-axis), as illustrated in Fig. 9(a) and the inset of Fig. 1.

Here, the region of interest for droplet dynamics is confined to $|z| \le d_{max}$ and is located sufficiently far from the magnetic poles, leading to significantly non-uniform field. Referring to Figs. 9(b) and (c) again, $O(B_r) \sim O(B_z)$, with $B_r$ exhibiting a significantly steeper radial increase. Moreover, as evident from Fig. 9(d), a marginal decrease in $B_z$ is noted for small values of *r* (i.e., in the vicinity of impact location), indicating a weaker $\partial B_z/\partial r$. Therefore, the horizontal magnetic force in the transverse direction can be approximated as,

$$F_T = F_r \approx \frac{\chi}{\mu_o}\left[B_r \frac{\partial B_r}{\partial r}\right] \sim \frac{\chi B_o^2}{\mu_o D_o} \tag{10}$$

Whereas, in the longitudinal direction it scales as,

$$F_L = F_z = \frac{\chi}{\mu_o}\left[B_r \frac{\partial B_r}{\partial z} + B_z \frac{\partial B_z}{\partial z}\right] \sim \frac{\chi B_o^2}{\mu_o D_o} \tag{11}$$

The expressions in Eqs. (10) and (11) reveal that the magnetic forces of scale $F_m \sim \chi B_o^2/\mu_o D_o$ act radially outward from the point of impact, thereby promoting overall droplet spreading. For analytical simplicity, the spreading enhancement induced by magnetic forces is assumed to be identical and uniform along all the radial directions in the spreading droplet, resulting in a cylindrical spreading lamella.

The change in spreading behaviour can be assessed by quantifying the magnetic work done $W_{m_s}$ during droplet spreading phase. The work done by magnetic forces in moving an infinitesimal magnetic fluid volume $dv$, from point of impact to a radial distance $R$ is expressed as,

$$dW_{m_s} = F_m dvR = \frac{\chi B_o^2}{\mu_o D_o} 2\pi R^2 dR H_l \tag{12}$$

Where, $H_l$ is the thickness of liquid lamella, which is assumed constant. Integrating Eq. (12) over complete spreading phase gives the total magnetic work during spreading,

$$W_{m_s} = \int_0^{R_{\max}} dW_{m_s} = \frac{2\pi \chi B_o^2 H_l R_{\max}^3}{3\mu_o D_o} \tag{13}$$

Where, $R_{\max} = D_{\max}/2$ is the maximum spread radius of the droplet. Assuming negligible mass loss, a conservation of volume between the pre-impact droplet and that at its maximum spread state yields

$$W_{m_s} = \frac{\chi B_o^2 D_{\max}}{3\mu_o D_o} \times \frac{1}{6}\pi D_o^3 \tag{14}$$

At the maximum spread state, the kinetic energy $KE_2$ of the droplet is zero and the surface energy $SE_2$ is given by [64]

$$SE_2 = \sigma\left[\frac{\pi}{4} D_{\max}^2 \left(1-\cos\theta\right) + \frac{2\pi}{3}\frac{D_o^3}{D_{\max}}\right] \times \frac{1}{6}\pi D_o^3 \tag{15}$$

Where, $\theta$ is the static contact angle of the ferrofluid droplet on the substrate studied. The energy loss through viscous dissipation during spreading is expressed by [65],

$$W_{VD} = \int_0^{t_s}\int_V \Phi dVdt \approx \Phi V t_s \tag{16}$$

Where, $\Phi$ is the viscous dissipation per unit volume per unit time, while $V$ represents the volume of the liquid layer where viscous losses occur; these quantities are given by:

$$\Phi \approx \eta\left(\frac{V_i}{\delta}\right)^2, \qquad V = \frac{\pi}{4}D_{\max}{}^2\delta \tag{17}$$

Here, $\delta = 2D_o/\sqrt{\mathrm{Re}}$ is the boundary layer thickness, where $\mathrm{Re} = \rho V_i D_o/\eta$ is the Reynolds number based on the effective droplet impact velocity, $V_i$. In line with the literature [66,67], the spreading time is scaled to be $t_s \approx D_{\max}/2V_i$. Therefore, Eq. (17) can be rewritten to obtain the expression of viscous dissipation:

$$W_{VD} = \frac{\pi}{16}\frac{\rho V_i^2 D_{\max}{}^3}{\sqrt{\mathrm{Re}}} \tag{18}$$

The energy balance between the droplet just before impact and at its maximum spread state can be expressed as:

$$KE_1 + SE_1 = KE_2 + W_{m_s} + W_{VD} + SE_2 \tag{19}$$

Combining this with Eqs. (7)-(18) yields,

$$\frac{1}{16}\frac{We_o}{\sqrt{\mathrm{Re}_o}}\left(1+\frac{Bo_m}{We_o}\right)^{\frac{3}{4}}\beta_{\max}{}^3 + \left(\frac{1-\cos\theta}{4}\right)\beta_{\max}{}^2 - \frac{1}{18}Bo_m\beta_{\max} + \frac{2}{3}\frac{1}{\beta_{\max}} - \frac{We_o}{12}\left(1+\frac{Bo_m}{We_o}\right) - 1 = 0 \tag{20}$$

Where, $We_o = \rho V_o{}^2 D_o/\sigma$ and $\mathrm{Re}_o = \rho V_o D_o/\eta$, are Weber number and Reynolds number based on the free fall impact velocity $V_o$ of droplets (corresponding to $Bo_m = 0$). Eq. (20) is solved to obtain the dimensionless maximum spread diameter (maximum spread factor) $\beta_{\max}$ of the ferrofluid droplet in presence of magnetic fields. The predicted values of $\beta_{\max}$ are plotted against $Bo_m$ (see Fig. 10), and are compared with the experimentally measured values. The model is found to predict both the $\beta_{\max}$ as well as its variation trend with $Bo_m$ reasonably well. In particular, for $We_o \approx 38$ ($h_o = 5$ cm), shown in Figs. 10(b) and (e), where the assumed linear magnetic field variation approximation is most accurate for our experimental conditions, the theoretical predictions show the closest agreement with the experimental data. For smaller release height $h_o = 3$ cm ($We_o \approx 21$), stronger magnetic fields do not decay completely to zero within the droplet traversal distance. Consequently, the average magnetic field gradients considered using linear decay approximation are larger (hence larger magnetic forces) than those observed experimentally.

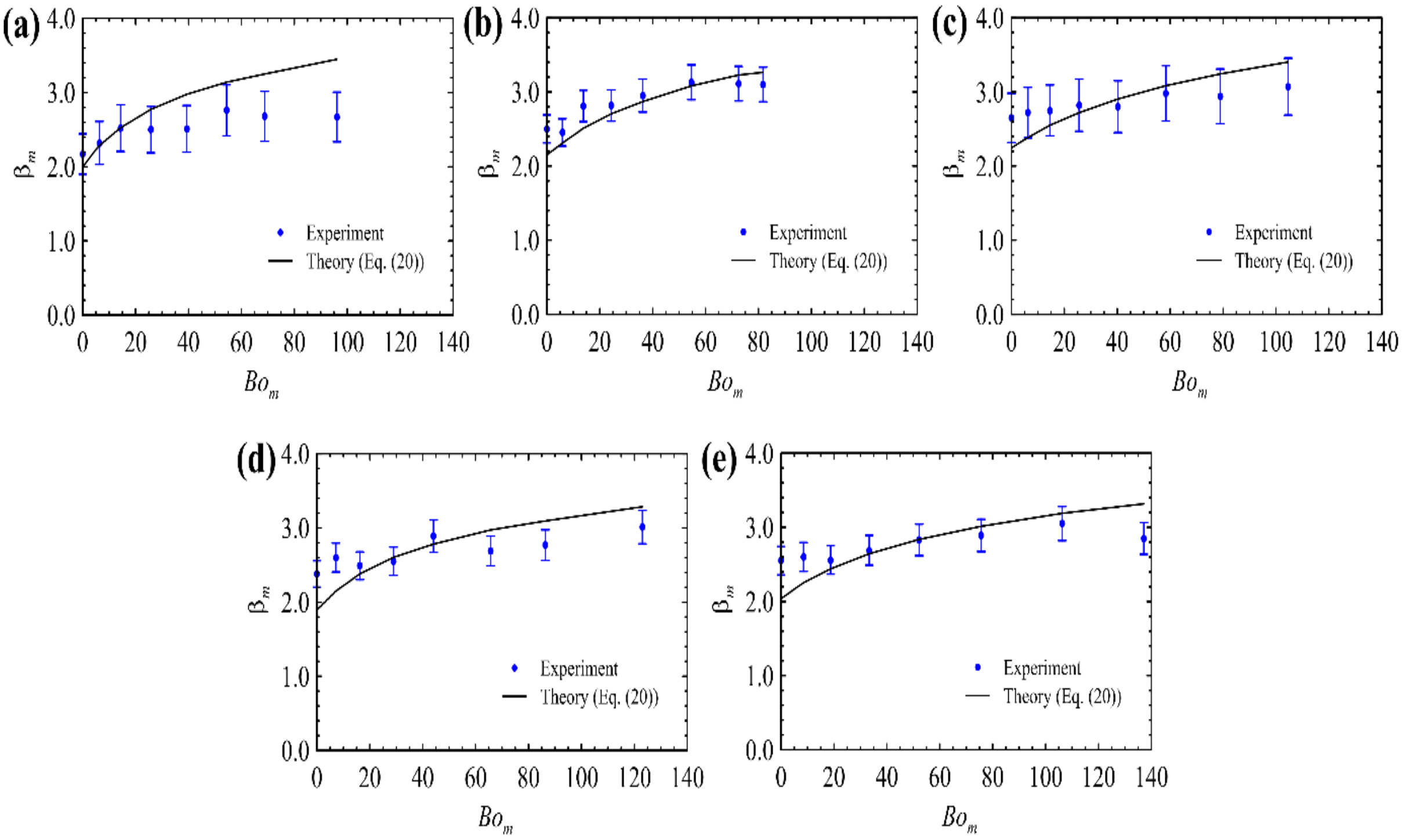

**Fig. 10.** Comparison between the predicted and experimentally measured maximum spread factor, $\beta_{\max}$, for (a) $We_o \approx 21$, 5% w/w, (b) $We_o \approx 38$, 5% w/w, (c) $We_o \approx 56$, 5% w/w, (d) $We_o \approx 21$, 7.5% w/w, and (e) $We_o \approx 38$, 7.5% w/w.

As a result, the model slightly overpredicts $\beta_{\max}$ at smaller release heights, particularly for higher $Bo_m$, as evident from Figs. 10(a) and (d), where magnetic fields decay over larger distances. In contrast, for smaller $Bo_m$, the magnetic field decays over shorter distances resulting in larger magnetic field gradients in reality, causing the model to slightly underpredict $\beta_{\max}$, which becomes more pronounced at higher release heights $h_o = 7$ cm ( $We_o \approx 56$). Nevertheless, besides providing insight into the subsequent magneto-Leidenfrost behaviour and associated hydrodynamics, the proposed theoretical model may also be useful for predicting the maximum spread diameter of ferrofluid droplets in magnetic field environments.

## 4. Conclusions

We investigated the impact dynamics and boiling behaviour of ferrofluid droplets on superheated substrates in the presence of magnetic field. Unlike previously reported approaches for Leidenfrost droplet manipulation [7,40], we employed a simple magnetic field-based technique to control ferrofluid impact outcome and subsequent boiling behaviour, and discovered a purely novel phenomenon of magnetic-field-induced Leidenfrost effect-driven droplet rebound, termed the MLFE. We discuss the underlying physics by scaling magnetic forces and their influence on ferrofluid droplets during different impact and spreading stages for a range of impact conditions, magnetic field strengths, and nanoparticle concentrations.

The findings reveal that sufficiently strong magnetic fields can induce droplet rebound at substrate temperatures below the zero-field dynamic Leidenfrost temperature ( $T_s < T_{DL}$ ) for a given $We_o$ and nanoparticle concentration. Magnetic forces enhance droplet spreading which increases heat transfer promoting a stable vapour layer formation that facilitates smooth recoil and rebound. The MLFE onset is characterized by a critical $Bo_m$, below which droplets remain attached to the substrate. At lower release height (lower $We_o$), a larger critical $Bo_m$ is required for droplet rebound. However, increasing the nanoparticle concentration enables rebound at even lower $We_o$, demonstrating the applicability of the technique even in space-constrained conditions.

In contrast to previous studies [58-61] reporting nearly constant droplet residence times in Leidenfrost regime $T_s \geq T_{DL}$, we observe a linear decrease in residence time (up to 50%) with $Bo_m$. At sufficiently high substrate temperatures ($T_s >> T_{DL}$), ferrofluid droplets exhibit a fragmenting Leidenfrost behaviour characterized by vigorous boiling, lamella rupture, and formation of filamentous structures. At low $Bo_m$, a major central liquid mass rebounds whereas high $Bo_m$ produces enhanced magnetic stretching and extensive lamella perforation resulting in complete fragmentation into much smaller daughter droplets.

Finally, we show that a non-uniform magnetic field generates both vertical and horizontal magnetic forces, enhancing droplet spreading through increased impact velocity and horizontal stretching. By incorporating these magnetic effects, we develop an energy-balance-based theoretical model to predict the maximum spread factor $\beta_{\max}$, the key parameter governing the magneto-Leidenfrost effect, with predictions in close agreement with experimental observations.

Our findings reveal a hitherto unknown phenomenon in colloids and interface science: the MLFE in ferrofluids, and offer a potential route for frictionless and rapid droplet transport, targeted heat-transfer modulation in thermal management applications, controlled reagent transport and deposition, magnetic manipulation of fluids, micro-droplets sorting, and allied technologies. The insights from this study expand the fundamental understanding of droplet impact phenomena in colloid and interface science under external forcing. The findings offer a novel external-stimulus-based means for enabling frictionless droplet transport [44,56,57] and thermal management applications, where control over impact outcome (rebounding/no rebounding) and residence time can enable targeted heat transfer manipulation. Further, the proposed critical $Bo_m$ criterion provides a guideline for tailoring magneto-Leidenfrost behaviour through appropriate combinations of ferrofluid composition and magnetic field strength, offering flexibility for applications having constraints over magnetic field strength or nanoparticle concentration.

**Data availability statement:** All data pertaining to the research is in the article.

**Conflict of interests:** The authors have no conflicts of interest regarding this research.

**Funding/Acknowledgement:** AKJ thanks the PMRF scheme, Govt. of India, for funding his PhD fellowship. NSB thanks the MoE, Govt. of India, for funding his PhD fellowship. PD thanks IIT Kharagpur and Anusandhan National Research Foundation (ANRF), India, for funding this research.

**CRediT authorship contribution statement:** ***AKJ*** – experiments, data processing, visualization, analysis, modeling, writing (initial and final draft); ***NSB*** – experiments, data processing, writing (review and editing); ***PD*** – conceptualization, visualization, resources, modeling, analysis, funding acquisition, supervision, writing (initial and final draft).